Prediction of properties of metal alloy materials based on machine learning

Houchen Zuo [a], Yongquan Jiang [b,*], Yan Yang [b], Jie Hu [a]

a.State key labratory of traction power, Southwest Jiaotong University, Chengdu 610031, China

b. School of Computing and Artificial Intelligence, Southwest Jiaotong University, Chengdu 611756, China

*. yqjiang@swjtu.edu.cn

Abstract

Density functional theory and its optimization algorithm are the main methods to calculate the properties in the field of materials. Although the calculation results are accurate, it costs a lot of time and money. In order to alleviate this problem, we intend to use machine learning to predict material properties. In this paper, we conduct experiments on atomic volume, atomic energy and atomic formation energy of metal alloys, using the open quantum material database. Through the traditional machine learning models, deep learning network and automated machine learning, we verify the feasibility of machine learning in material property prediction. The experimental results show that the machine learning can predict the material properties accurately.

1. Introduction

At present, there are two main trends in the field of material properties calculation. One is to build mathematical models based on professional knowledge and experience. For example, density functional theory can predict the properties of compounds in small-scale and medium-scale systems and provide high accuracy with reasonable calculation cost. However, for large-scale systems, the traditional methods of building mathematical models either perform poorly or can only be at the cost of huge computing costs. The most important thing is that these methods require professional knowledge and experience. The other is based on the material database. This kind of methods benefits from the hot development of machine learning. The applications of machine learning in the field of material prediction are growing rapidly and the prediction results can reach the calculation accuracy of quantum mechanics software [1], [2], [3]. Most importantly, machine learning can reduce the prior knowledge required for entry. Through machine learning, given enough data and algorithms, the computer can discover known rules without human intervention, and even discover potential and unknown rules. In this paper, we use several representative methods of machine learning to predict material properties.

2. Data and methods

2.1 Data extraction

We use open quantum material database (OQMD) [4], [5], including the data of about 300,000 compounds calculated by density functional theory. Metal compound is an important phase in many alloys, which plays an important role in strengthening, so we use the alloy data containing only metal elements to experiment, which contain repeated compounds, but the pressure conditions are different. We choose average atomic volume (volume per atom), average atomic energy (energy per atom) and atomic formation energy for experiments. Atomic volume is often used to estimate the space occupied by an atom in the crystal. Atomic energy is mainly the kinetic energy and potential energy of the electrons outside the nucleus. Atomic formation energy is an important index to evaluate the stability of crystal structure. The feature set we constructed uses element composition (atomic frequency) and pressure in six directions as input characteristics. Only atomic components used as input features can't reflect the fact that the microstructure and properties of compounds with the same composition may be different under different pressure conditions. In order to solve this problem, we extracted the atomic frequency of 47 metal elements and the pressure in six directions, a total of 53 dimensions as input features. The output features include average atomic volume, average atomic energy and atomic formation energy. There are 6135 records of average atomic volume and average atomic energy, including 2596 compounds, and 10017 records of atomic formation energy, including 2596 compounds. We divide the dataset, 90% of which are training set and 10% are test set.

2.2 Methods and evaluation standards

We use two traditional machine learning models. The first model is support vector machine (SVM). The second model is gradient boosting regression (GBR). Because GBR uses a set of weak learners, there are many super parameters, so grid search technology is used in the experiment. We also build two deep neural network models. The first model is composed of fully connected layer, referred to as DNN. The second model adds residual block [6] on the basis of the first model, referred to as ResNet. We also use batch normalization [7] and dropout [8] strategies in the network. Finally, we use the automated machine learning [9], [10], which purpose is to automatically explore and construct relevant features, select the most appropriate model, set its optimal parameters, and select an optimization algorithm. In this paper, we use AutoKeras [11] and auto_ml [12] to verify the feasibility of automated machine learning. AutoKeras and auto_ml are frameworks.

Regression problems does not involve the concept of accuracy, so coefficient of determination R-square is used as the evaluation standard, and its value range is generally between 0 and 1. The closer to 1, the better the model effect is. However, if the fitting effect is worse than the average value of all, R-square will be negative. We use it as the evaluation standard, and the formula is shown in (1). In addition, we also use MAE to evaluate the experimental model. MAE is the average of the absolute error between the predicted value and the real value, and the formula is shown in (2). The code of relevant work has been uploaded to github.com/ahzzhc/oqmd-nn.

$$R-square: R^2 = 1 - \frac{\sum_{i=1}^{n} w_i(\hat{y}_i - \bar{y}_i)^2}{\sum_{i=1}^{n} w_i(y_i - \bar{y}_i)^2} \quad (1)$$

$$MAE: C = \frac{1}{n}\sum_{i=1}^{n}|y_i - \hat{y}_i| \quad (2)$$

In formulas, n is number of samples, $w_i$ is weight coefficient, $\bar{y}_i$ is average value, $y_i$ is real data, $\hat{y}_i$ is forecast data.

3. Results and discussions

We use SVM, GBR, DNN, ResNet, auto_ml and AutoKeras to predict atomic formation energy. The experimental results are shown in Fig. 1. We also use average atomic volume dataset to further verify the feasibility of machine learning, and the results are shown in Fig. 2. The closer the data is to the diagonal, the more accurate the data prediction is. Most of the data of atomic formation energy and average atomic volume are distributed on both sides of the diagonal. It can be seen that machine learning can predict these attributes more accurately. From the loss diagram, it can be seen that AutoKeras can also achieve good results.

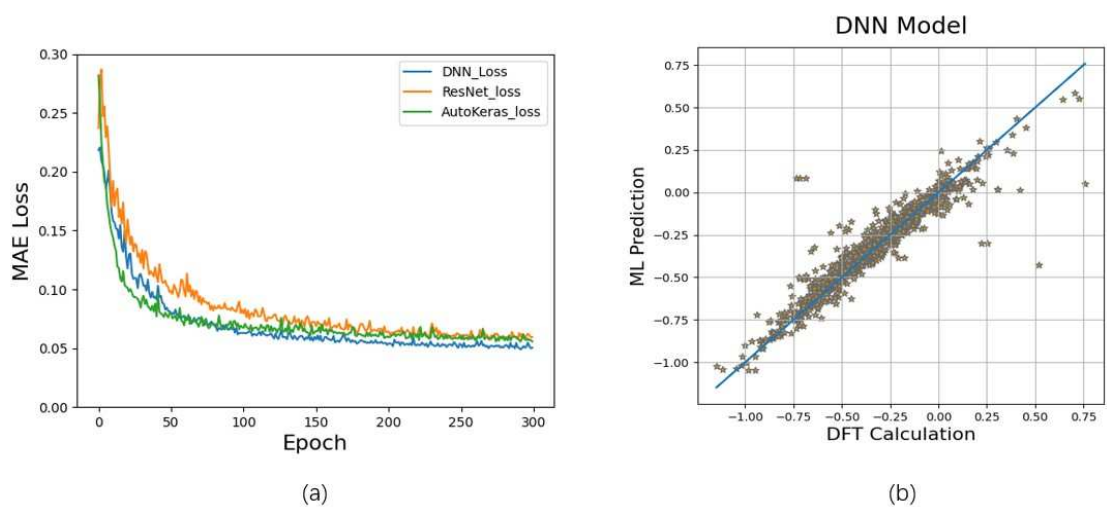

(a)          (b)

Fig. 1 (a) The loss curve of DNN, ResNet and AutoKeras in the test set. (b) The best model for predicting atomic formation energy, the abscissa is the accurate result of DFT calculation, and the ordinate is the prediction result of the model.

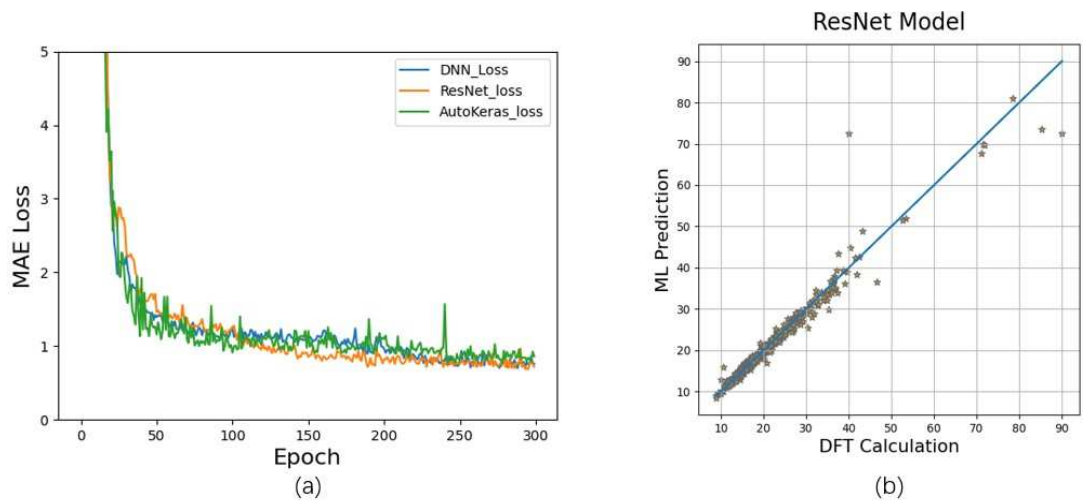

Fig. 2 (a) The loss curve of DNN, ResNet and AutoKeras in the test set. (b) The best model for predicting average atomic volume.

We study the best models provided by auto_ml and AutoKeras, and find that the models provided by auto_ml have no change. We study the source code of auto_ml and find that layers and neurons of the deep learning model provided by auto_ml have been determined. The default layer number is 4, and the neurons of each layer depend on the dimension of input features. In each layer, the input dimension multiplies a coefficient, and the result is compared with 10. The minimum value is taken as the number of neurons. The coefficients of the first three layers are 1, 0.75, 0.25. The best model provided by auto_ml is not very meaningful, but we can optimize the model based on the best model provided by AutoKeras.

We use dataset of average atomic energy for experiment. This time, we first use AutoKeras for prediction, and then build models according to the best model provided by AutoKeras. The experimental results are shown in Fig. 3. Based on the network structure and parameters provided by AutoKeras, we can build better models. We also try multi-attribute prediction. The data number of atomic formation energy is different from the other two properties, so we use average atomic volume and average atomic energy. We build the model based on the best model provided by AutoKeras, and find that multi-attribute prediction can not only achieve good results but also save running time. We summarize the experimental results as shown in Fig. 4.

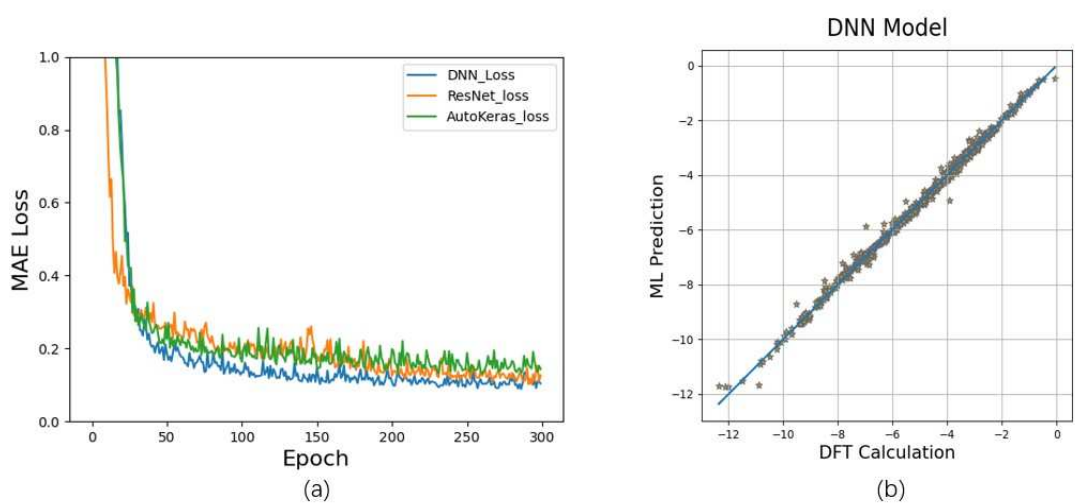

Fig. 3 (a) The loss curve of DNN, ResNet and AutoKeras in the test set. (b) The best model for predicting average atomic energy.

| Model | Atomic Formation Energy | | | Average Atomic Energy | | | Average Atomic Volume | | |
|---|---|---|---|---|---|---|---|---|---|
| | MAE(eV /atom) | R-square | Time(s) | MAE(eV /atom) | R-square | Time(s) | MAE($10^{-30}$ m$^3$/atom) | R-square | Time(s) |
| SVM | 0.116 | 0.666 | 2.5 | 0.647 | 0.756 | 1.5 | 2.904 | 0.440 | 1.8 |
| GBR | 0.090 | 0.806 | 225.3 | 0.406 | 0.938 | 135.8 | 2.105 | 0.774 | 132.7 |
| DNN | 0.051 | 0.891 | 39.9 | 0.105 | 0.995 | 27.3 | 0.761 | 0.935 | 26.8 |
| ResNet | 0.060 | 0.866 | 27.4 | 0.127 | 0.991 | 27.9 | 0.721 | 0.956 | 21.5 |
| auto_ml | 0.169 | 0.386 | 174.0 | 0.167 | 0.989 | 168.0 | 0.907 | 0.950 | 121.0 |
| AutoKeras | 0.056 | 0.880 | 18.3 | 0.144 | 0.939 | 9.1 | 0.861 | 0.939 | 17.6 |
| DNN-multiple | | | | 0.204 | 0.981 | 26.6 | 0.648 | 0.970 | 26.6 |

Fig. 4 Experimental results of different models, the calculation time is based on rtx2060.

4. Conclusions

The development of machine learning in the field of materials can solve the time and financial consumption of traditional mathematical model. With the continuous improvement of the accuracy of the prediction results, the prediction algorithm can replace the experiment to a certain extent. We apply machine learning to predict single attribute and multi-attribute by using average atomic volume, average atomic energy and atomic formation energy. Through experiments, we find that machine learning can achieve very good results. We also use automated machine learning models in the experiment. The experimental results show that these models can not only achieve good results, but also save the time of parameter adjustment. We can also improve the models based on the best model provided by automated machine learning to get better models.